\documentclass[12pt]{article}
\usepackage[numbers]{natbib}
\usepackage{a4}
\usepackage{amsmath}
\usepackage{amssymb}
\usepackage{latexsym}
\usepackage{amssymb}
\usepackage{epsfig}
\usepackage{natbib}
\def \ii{{\mathrm{i}}}
\def \TT{{\mathrm{T}}}
\def \TL{{\mathrm{L}}}

\def \d{{\mathrm{d}}}

\def \R{{\mathbb{R}}}

\def \pd{\partial}

\def \e{{\mathrm{e}}}

\def \Bsigma{\boldsymbol{\sigma}}
\def \Bbeta{\boldsymbol{\beta}}

\def \Bs{\boldsymbol{s}}

\def \Bu{{\boldsymbol{u}}}

\def \Bv{{\boldsymbol{v}}}
\def \BR{{\boldsymbol{R}}}
\def \BV{{\boldsymbol{V}}}
\def \BQ{{\boldsymbol{Q}}}
\def \BF{{\boldsymbol{F}}}

\def \BJ{{\boldsymbol{J}}}

\def \Bbeta{\boldsymbol{\beta}}

\def \rr{{\boldsymbol{r}}}
\def \RR{{\boldsymbol{R}}}

\def \Bu{{\boldsymbol{u}}}
\def \Bv{{\boldsymbol{v}}}

\def \Bp{{\boldsymbol{p}}}
\def \BJ{{\boldsymbol{J}}}
\def \BA{{\boldsymbol{A}}}
\def \BB{{\boldsymbol{B}}}
\def \BE{{\boldsymbol{E}}}

\def \Bu{{\boldsymbol{u}}}
\def \Bv{{\boldsymbol{v}}}

\def \Bp{{\boldsymbol{p}}}
\def \BB{{\boldsymbol{B}}}
\def \BE{{\boldsymbol{E}}}
\def \BH{{\boldsymbol{H}}}
\def \BD{{\boldsymbol{D}}}
\def \VV{{\cal{V}}} 
\def \tr{{t_{\text{r}}}}

\def \Bz{{\boldsymbol{z}}}

\begin{document}
\title{{\bf On retardation, radiation and Li\'enard-Wiechert type 
potentials 
in electrodynamics and elastodynamics
}}
\author{
Markus Lazar~$^\text{}$\footnote{
{\it E-mail address:} lazar@fkp.tu-darmstadt.de (M.~Lazar).
\newline
Tel.:~+49(0)6151/163686;~Fax.:~+49(0)6151/163681.
}
\\ \\
${}^\text{}$ 
        Heisenberg Research Group,\\
        Department of Physics,\\
        Darmstadt University of Technology,\\
        Hochschulstr. 6,\\      
        D-64289 Darmstadt, Germany\\
}

\date{\today}    
\maketitle

\begin{abstract}
The aim of this paper is to investigate 
the fundamental problems of retardation and radiation 
caused by non-uniformly moving point sources 
using the theories of electrodynamics and elastodynamics.
This paper investigates and compares
the retarded electromagnetic fields and the retarded elastodynamic fields.
For the non-uniform motion 
of a general point source, the Li\'enard-Wiechert type 
potentials and the radiation and nonradiation fields are derived
for a point charge and for a point force. 
\\

\noindent
{\bf Keywords:} non-uniform motion; electrodynamics; 
elastodynamics; radiation; retardation.\\
\end{abstract}
\newpage

\section{Introduction}
A lot of research has been devoted to problems in 
electrodynamics and elastodynamics.
Often the investigations use some analogies, the main 
differences are due to the vector character of the electromagnetic fields
and the tensor character of elastodynamic fields.
Other important differences are that electrodynamics possesses only
one characteristic velocity (velocity of light), and the fundamental solution
of the field equations is a scalar-valued generalized function.
On the other hand, elastodynamics of isotropic materials 
has two characteristic velocities (longitudinal and transversal velocity of the
elastic waves) and the fundamental solution of the field equation 
is a tensor-valued generalized function (see Table~1). 
These differences may be significant for the explicite expressions,
but often the analysis is very similar from the mathematical point of
view.  
However, the dynamics of elastodynamic waves is more complicated than 
the dynamics of electromagnetic waves.

In this paper we compare the fundamentals and waves 
in electrodynamics and elastodynamics, and 
this research is motivated by analogies.
We review the retarded causal fields 
and the non-uniform motion of a point charge in  electrodynamics. 
Motivated by straightforward analogies, 
we have investigated the
retarded fields and the non-uniform motion of a point force in elastodynamics.
\begin{table}[b]
\caption{Comparison between electrodynamics and elastodynamics}
\begin{center}
\leavevmode
\begin{tabular}{||c|c||}\hline
Electrodynamics & Elastodynamics\\
\hline
Green function $G(r,t)$  & Green tensor $G_{ij}(r,t)$\\
1 velocity of light: $c$ & 2 velocities of sound: $c_\TT$, $c_\TL$ \\
1 retarded time: $\tr=t-R/c$ & 2 retarded times: $t_{\TT}=t-R/c_{\TT}$\\
 &\qquad\qquad\qquad\quad\ \   $t_{\TL}=t-R/c_{\TL}$\\
\hline
\end{tabular}
\end{center}
\label{table}
\end{table}

In this paper we give the answers to the fundamental questions:
\begin{itemize}
\item
What is the elastodynamic Li\'enard-Wiechert potential of a point force?
\item
What are the equations in analogy to the  Li\'enard-Wiechert form 
and the Heaviside-Schott-Feynman form of the electromagnetic field strengths?
\item
What is the form of the elastodynamic radiation fields?
\end{itemize}

One important property of the dynamics of waves 
is the retardation, which is a 
consequence of the finite speed of the propagating fields.
There is always a time delay, 
since an effect observed by the receiver at the present position 
and present time was caused by the sender at some earlier time (retarded time) 
and at the retarded position.

This paper is organized as follows.
In Section~2, we review the theory of electrodynamics.
First of all, the retarded fields are presented. Especially, the Jefimenko
equations are presented as the retarded electromagnetic field
strengths, which are the causal solutions of the Maxwell equations.
Later, we consider the non-uniform motion of a point charge with time-dependent
magnitude. The so-called Heaviside-Schott-Feynman equations will be discussed
as well as the Li\'enard-Wiechert form of the electromagnetic field strengths.
The electromagnetic radiation produced by the point charge will be given 
explicitly.
In Section~3, we consider the theory of elastodynamics and
we investigate the retarded displacement, velocity and distortion fields.
The non-uniform motion of a point force is studied. We calculate the
elastodynamic radiation and non-radiation parts.
We show that the fields of a non-uniform moving point force are the
generalization of the Stokes solution towards non-uniform motion.
Also we consider the case of time-harmonic forces.
All the mathematical expressions concerning the derivative with respect 
to the retarded time are given in an Appendix.

\section{Electrodynamics}
In this section we investigate the retarded fields and the radiation 
of electromagnetic fields using the theory of electrodynamics.

\subsection{Basic framework and retarded fields}

The basic electromagnetic field laws are represented by the 
inhomogeneous and homogeneous Maxwell equations~\citep{Jackson,Griffiths,CD}
\begin{align}
\label{ME-inh}
&\nabla\cdot \BD=\rho\,,\qquad \nabla\times\BH-\pd_t\BD=\BJ\,,\\
\label{ME-h}
&\nabla\cdot \BB=0\,,\qquad \nabla\times\BE+\pd_t\BB=0\,,
\end{align}
where $\BD$ is the electric displacement vector (electric excitation), 
$\BH$ is the magnetic excitation vector, 
$\BB$ is the magnetic field strength vector, $\BE$ is the electric field
strength vector, $\BJ$ is the electric current density vector, and 
$\rho$ is the electric charge density.
In addition, the electric current density vector 
and the electric charge density
fulfill the continuity equation 
\begin{align}
\label{CE}
&\nabla\cdot \BJ+\pd_t\rho=0\,.
\end{align}
The constitutive equations for the fields in a vacuum read
\begin{align}
\label{CE2}
\BD=\epsilon_0\, \BE\,,\qquad 
\BH=\frac{1}{\mu_0}\, \BB\,,
\end{align}
where $\epsilon_0$ is the vacuum permittivity and 
$\mu_0$ is the vacuum permeability.
The speed of light in vacuum is given by
\begin{align}
c^2=\frac{1}{\epsilon_0\mu_0}\,.
\end{align}

The electromagnetic field strengths can be expressed in terms
of the electromagnetic gauge potentials $\phi$ and $\BA$
(scalar potential $\phi$ and vector potential $\BA$):
\begin{align} 
\label{E}
\BE&=-\nabla \phi-\pd_t \BA\,,\\
\label{B}
\BB&=\nabla\times \BA\,.
\end{align}
Using the Lorentz gauge condition: 
\begin{align}
\label{LG}
\frac{1}{c^2}\,\pd_t \phi+ \nabla \cdot \BA=0\,,
\end{align}
the electromagnetic gauge potentials fulfill the following 
inhomogeneous wave equations
\begin{align}
\label{phi-w}
\square\,\phi=\frac{1}{\epsilon_0}\, \rho
\end{align}
and
\begin{align}
\label{A-w}
\square\,\BA=\mu_0\, \BJ\,,
\end{align}
where the d'Alembert operator is defined by
\begin{align}
\square:=\frac{1}{c^2}\,\pd_{tt}-\Delta\,\qquad\text{with}\quad
\Delta=\nabla\cdot\nabla\ .
\end{align}

For zero initial conditions, 
the solutions of Eqs.~(\ref{phi-w}) and (\ref{A-w}) may be written as
space-time convolution integrals
\begin{align}
\phi(\rr,t)=
\label{phi-C}
\frac{1}{\epsilon_0}\int_{-\infty}^t \int_{-\infty}^\infty
G(\rr-\rr', t-t')\, \rho(\rr',t')\, \d \rr'\, \d t'\, \\
\BA(\rr,t)=
\label{A-C}
\mu_0 \int_{-\infty}^t \int_{-\infty}^\infty
G(\rr-\rr', t-t')\, \BJ(\rr',t')\, \d \rr'\, \d t'\, .
\end{align}
Using the 3D Green function of the wave equation (e.g.~\citep{Barton})
\begin{align}
\label{GF}
G(\rr,t)=\frac{1}{4\pi r}\, \delta(t-r/c)\,,
\end{align}
we obtain the retarded electromagnetic potentials, which were originally 
introduced by~\citet{Lorenz}, and they read~\citep{Jackson,Griffiths}:
\begin{align}
\label{phi-ret}
\phi(\rr,t)&=
\frac{1}{4\pi \epsilon_0}
\int_\VV \frac{\rho(\rr',t-R/c)}{R}\, \d \rr'\,,\\
\label{A-ret}
\BA(\rr,t)&=
\frac{1}{4\pi \epsilon_0 c^2 }
\int_\VV \frac{\BJ(\rr',t-R/c)}{R}\, \d \rr'\,,
\end{align}
where $\BR=\rr-\rr'$, $R=|\rr-\rr'|$, $\rr\in\R^3$, $t\in\R$ 
and $\VV$ denotes the whole three-dimensional space.
The idea of a retarded scalar potential was first developed by \citet{Lorenz61} 
in 1861 while studying waves in the theory of elasticity.
The retarded potentials fulfill the 
Lorentz gauge condition (see, e.g.,~\citep{Griffiths,Jefimenko}).

Substituting the retarded potentials~(\ref{phi-ret}) and (\ref{A-ret}) 
into the definition of the electromagnetic field strengths~(\ref{E}) and 
(\ref{B}), and using the relations 
\begin{align}
&\nabla \rho(\rr',t-R/c)=-\frac{\BR}{c R}\, \pd_t
 \rho(\rr',t-R/c)\,,\quad
&\nabla \BJ(\rr',t-R/c)=-\frac{\BR}{c R}\, \pd_t
 \BJ(\rr',t-R/c)\,,
\end{align}
we obtain the retarded electromagnetic field
strength vectors:
\begin{align}
\label{E-J}
\BE(\rr,t)&=
\frac{1}{4\pi \epsilon_0}
\int_\VV \bigg(
\frac{\rho(\rr',t-R/c)}{R^3}\,\RR
+\frac{\pd_t \rho(\rr',t-R/c)}{c R^2}\,\RR
-\frac{\pd_t \BJ(\rr',t-R/c)}{c^2 R}\bigg)
 \d \rr'\,,\\
\label{B-J}
\BB(\rr,t)&=
\frac{\mu_0}{4\pi}
\int_\VV \bigg(
\frac{\BJ(\rr',t-R/c)}{R^3}
+\frac{\pd_t \BJ(\rr',t-R/c)}{c R^2}\bigg)\times \RR\ 
 \d \rr'\,.
\end{align}
Eq.~(\ref{E-J}) is the time-dependent generalized Coulomb-Faraday law
and Eq.~(\ref{B-J}) is the time-dependent generalized Biot-Savart law 
(see also \citep{HM}).
Eqs.~(\ref{E-J}) and (\ref{B-J}) express the electromagnetic fields in
terms of their retarded sources $\rho$, $\BJ$, $\pd_t\rho$ and $\pd_t \BJ$
with full generality. 
They were originally derived by~\citet{Jefimenko} (see also~\citep{Jefimenko04}).
They also appear in the book of~\citet{CD} and in the third edition
of \citet{Lorrain}.
An equivalent representation was given by~\citet{PP} 
(see also~\citep{Jefimenko04}).
Nowadays both equations are called the Jefimenko equations 
in standard books on electrodynamics (e.g.~\citep{Jackson,Griffiths,HM}). 
They are fundamental, elegant, 
and very useful equations.

\subsection{A non-uniformly moving point charge}
Now we consider 
a non-uniformly moving point charge carrying the time-dependent charge $q(t)$ 
at the position $\Bs(t)$.
The electric charge density and the electric current density vector are
given by
\begin{align}
\label{J}
\rho(\rr,t)&=q(t)\,\delta(\rr-\Bs(t))\,,\qquad
\BJ(\rr,t)=q(t)\BV(t)\, \delta(\rr-\Bs(t))\,,
\end{align}
where $\BV(t)=\pd_t{\Bs}(t)=\dot{\Bs}(t)$ 
is the arbitrary velocity of the non-uniform motion.
We consider the case that the velocity of the point charge
is less than the speed of light: $|\BV|<c$. 
Substitution of Eq.~(\ref{J}) in Eqs.~(\ref{phi-ret}) and (\ref{A-ret}) 
gives
\begin{align}
\label{phi-LW0}
\phi(\rr,t)&=
\frac{1}{4\pi\epsilon_0}
\int_\VV q(t-R/c)\, \frac{\delta(\rr'-\Bs(t-R/c))}{R}\, \d \rr'
\, ,\\
\label{A-LW0}
\BA(\rr,t)&=
\frac{1}{4\pi\epsilon_0 c^2}
\int_\VV q(t-R/c) \BV(t-R/c)\, \frac{\delta(\rr'-\Bs(t-R/c))}{R}\, \d \rr'
\, ,
\end{align}
Here remain the integrals of the delta functions, which are done by changing 
the variable of integration from $\rr'$ to $\Bz=\rr'-\Bs(t-R/c)$ 
with (see, e.g., \citep{Jones,Jones86,Eyges})
\begin{align}
\label{dz}
\d\Bz=J\, \d\rr'
\end{align}
and using the Jacobian $J$ of this transformation 
\begin{align}
\label{Jac}
J={\text{det}}\, \bigg(\frac{\pd\Bz}{\pd\rr'}\bigg)
=1-\frac{\BV(t-R/c)\cdot(\rr-\rr')}{c|\rr-\rr'|}\,,
\end{align}
we obtain
\begin{align}
\label{int-J}
\int 
F(\rr')\,\delta(\rr'-\Bs(t-R/c))\, \d \rr'&=
\int F(\rr')\,\delta(\Bz)\,\frac{1}{J}\, \d \Bz\nonumber\\
&=\frac{F(\rr')}{J}\bigg|_{\Bz=0}
=\frac{F(\rr')}{1-\frac{\BV(t-R/c)\cdot(\rr-\rr')}{c|\rr-\rr'|}}
\bigg|_{\rr'=\Bs(\tr)}\,.
\end{align}
From the argument of the $\delta$-function we get $\rr'=\Bs(\tr)$. 
Therefore, now $\RR(\tr)=\rr-\Bs(\tr)$ and 
$R(\tr)=|\rr-\Bs(\tr)|$ are time-dependent
and the retarded time is now given by
\begin{align}
\label{tr2}
\tr=t-|\rr-\Bs(\tr)|/c=t-R(\tr)/c  \,,
\end{align}
where $\Bs(\tr)$ is the retarded position of the moving source point.
For a point charge moving with velocity less than the speed of light in
free space ($|\BV|<c$), there is only one retarded time $\tr$ that 
satisfies Eq.~(\ref{tr2}) for each time $t$. 
In addition, we define
\begin{align}
\label{P}
P(t')=R(t')-\BV(t')\cdot\BR(t')/c\,, 
\end{align}
after the integration in $\rr'$, Eqs.~(\ref{phi-LW0}) 
and (\ref{A-LW0}) transform into
the Li\'enard-Wiechert potentials 
(scalar potential $\phi$ and vector potential $\BA$) 
of a point charge~\citep{Griffiths,Sommerfeld}:
\begin{align}
\label{phi-LW}
\phi(\rr,t)=
\frac{1}{4\pi \epsilon_0}
\bigg[\frac{q(t')}{P(t')}\bigg]_{t'=\tr},\qquad
\BA(\rr,t)=
\frac{1}{4\pi \epsilon_0 c^2 }
\bigg[\frac{q(t')\BV(t')}{P(t')}\bigg]_{t'=\tr}\,.
\end{align}
These are the Li\'enard-Wiechert potentials~\citep{Lienard,Wiechert} 
for a point charge with time-dependent magnitude.
Of course, the Li\'enard-Wiechert potentials 
fulfill the Lorentz gauge condition
(see, e.g., \citep{Jackson,LL}).

Substituting Eq.~(\ref{J}) into the Jefimenko equations~(\ref{E-J}) 
and (\ref{B-J}) gives the so-called Heaviside-Schott-Feynman 
formulae~\citep{Heaviside,Schott,Feynman} 
for the electric field strength $\BE$ and the magnetic field strength $\BB$ 
of a non-uniformly moving point charge~\citep{Jackson,HM}:
\begin{align}
\label{E-HF}
\BE(\rr,t)&=
\frac{1}{4\pi \epsilon_0}
\bigg(
\bigg[\frac{q(t') \RR(t')}{R^2(t') P(t')}\bigg]_{t'=\tr}
+\frac{1}{c}\,\pd_t 
\bigg[\frac{q(t')\RR(t')}{R(t') P(t')}\bigg]_{t'=\tr}
-\frac{1}{c^2}\,\pd_t 
\bigg[\frac{q(t')\BV(t')}{P(t')}\bigg]_{t'=\tr}
\bigg)\,,\\
\label{B-HF}
\BB(\rr,t)&=
\frac{1}{4\pi \epsilon_0 c^2}
\bigg(
\bigg[\frac{q(t')\BV(t')\times \RR(t')}{R^2(t') P(t')}\bigg]_{t'=\tr}
+\frac{1}{c}\,\pd_t 
\bigg[\frac{q(t')\BV(t')\times\RR(t')}{R(t') P(t')}\bigg]_{t'=\tr}
\bigg)\,.
\end{align}

We continue with the 
Li\'enard-Wiechert form of the electromagnetic field strengths
for the point charge with time-dependent magnitude.
Since the time derivative in Eqs.~(\ref{E-HF}) and 
(\ref{B-HF}) is involved, it is clear that the electromagnetic
fields~(\ref{E-HF}) and (\ref{B-HF}) 
will be functions not only of the velocity~$\BV$,
but also of the acceleration~$\dot{\BV}$, and of the time-derivative of
the magnitude of the point charge $\dot{q}$.  
We may therefore separate $\BE$ and $\BB$ into two parts each, one of
which involves the acceleration and the time-derivative of the charge 
and goes to zero for $\dot{\BV}=0$ and $\dot{q}=0$,
and one of which involves only the velocity $\BV$:
\begin{align}
\BE(\rr,t)&=\BE^{\text{nonr}}(\rr,t)+\BE^{\text{rad}}(\rr,t)\,,\\
\BB(\rr,t)&=\BB^{\text{nonr}}(\rr,t)+\BB^{\text{rad}}(\rr,t)\,.
\end{align}
The fields 
$\BE^{\text{nonr}}$ and $\BB^{\text{nonr}}$ are called 
the nonradiation or velocity parts of the electromagnetic field strengths, 
and the fields $\BE^{\text{rad}}$ and $\BB^{\text{rad}}$ 
are called the radiation or acceleration parts.
To obtain the Li\'enard-Wiechert form of the electromagnetic fields,
we must carry out the time-derivatives, which are not trivial because of the
subtle relation between the present and retarded time. 
Using Eqs.~(\ref{dt-Q})--(\ref{dt-PR}), 
the result reads 
\begin{align}
\BE^{\text{nonr}}(\rr,t)&=
\label{E-V}
\frac{1}{4\pi \epsilon_0}
\bigg[
\frac{q(t')}{P^3(t')}
\bigg(1-\frac{V^2(t')}{c^2}\bigg)
\bigg(\RR(t')-\frac{\BV(t')R(t')}{c}\bigg)\bigg]_{t'=\tr}\,,
\end{align}
\begin{align}
\BE^{\text{rad}}(\rr,t)&=
\frac{1}{4\pi \epsilon_0}
\bigg[
\bigg(\frac{\dot{q}(t')}{c\, P^2(t')} 
+\frac{q(t') \dot\BV(t')\cdot\RR(t')}{c^2P^3(t')}\bigg)
\bigg(\BR(t')-\frac{\BV(t') R(t')}{c}\bigg)
-\frac{q(t')\dot\BV(t') R(t')}{c^2 P^2(t')}\bigg]_{t'=\tr}\,,
\end{align}
\begin{align}
\BB^{\text{nonr}}(\rr,t)&=
\frac{1}{4\pi \epsilon_0 c^2}
\bigg[
\bigg(1-\frac{V^2(t')}{c^2}\bigg)
\frac{q(t') \BV(t')\times\RR(t')}{P^3(t')}\bigg]_{t'=\tr}
\end{align}
and
\begin{align}
\BB^{\text{rad}}(\rr,t)&=
\label{B-A}
\frac{1}{4\pi \epsilon_0c^2}
\bigg[\bigg(\frac{\dot{q}(t')}{c\, P^2(t')} 
+\frac{q(t') \dot\BV(t')\cdot\RR(t')}{c^2P^3(t')}\bigg)
\BV(t')\times \RR(t') 
+\frac{q(t') \dot\BV(t')\times \RR(t')}{c\, P^2(t')}\bigg]_{t'=\tr}\,.
\end{align}
It holds: $\BB=\RR\times \BE/cR$. 
We can see a clear separation into the near field or non-radiation field 
(velocity-dependent, 
which falls off as $1/R^2$) and the far field or radiation field 
(which falls off as $1/R$).
The nonradiation fields are identical with the `convective'
fields of a uniformly moving charge (see, e.g.,~\citep{PP}).
If the charge is accelerated, the electromagnetic fields are neither
static nor convective, and there is a net change in the field energy which
causes radiation. 
The electromagnetic radiation possesses two sources, and 
this is caused by the time-change of the 
charge $\dot{q}$ and the acceleration $\dot\BV$. 
For a constant charge $\dot{q}=0$, we recover in Eqs.~(\ref{E-V})--(\ref{B-A})
the original Li\'enard-Wiechert form of the electromagnetic fields
(see also~\citep{CD,HM,Jefimenko04,MH,Smith}).

For a localized point charge with time-dependent charge we obtain
\begin{align}
\label{E-q}
\BE(\rr,t)=\frac{1}{4\pi\epsilon_0} \frac{\BR}{R^3}\Big[
q(t-R/c)+\frac{R}{c}\, \dot{q}(t-R/c)\Big]
\end{align}
and $\BB=0$.

\section{Elastodynamics}
In this section we investigate the retarded fields and the radiation 
of elastodynamic fields using the theory of elastodynamics.

\subsection{Basic framework and retarded fields}
In elastodynamics, the force equilibrium condition reads 
(e.g. \citep{Achenbach,Miklowitz})
\begin{align}
\label{EC0}
\pd_t {p}_i -\pd_j \sigma_{ij}=F_i\,,
\end{align}
where $\Bp$, $\Bsigma$, and $\BF$ are the linear momentum vector, the force
stress tensor, and the body force vector. 
The constitutive relations are
\begin{align}
\label{CR-p}
p_i&= \rho\,  v_i=\rho\, \pd_t{u}_i\,,\\
\label{CR-t}
\sigma_{ij}&=C_{ijkl}\, \beta_{kl}=C_{ijkl}\, \pd_l u_{k}\,,
\end{align}
where $\Bv=\pd_t \Bu$ is the velocity vector of the continuum (particle velocity),
$\Bbeta=(\nabla \Bu)^T$ is the distortion tensor (displacement gradient), 
and $\Bu$ denotes the displacement vector.
Here $\rho$ denotes the mass density, and 
$C_{ijkl}$ is the tensor of elastic moduli.
If we substitute the constitutive relations~(\ref{CR-p}) and (\ref{CR-t}) in 
Eq.~(\ref{EC0}), we obtain an inhomogeneous Navier equation for 
the displacement vector $\Bu$ 
\begin{align}
\label{Navier}
\rho\, \pd_{tt}{u}_i -C_{ijkl}\pd_j\pd_l u_{k}=F_i\, .
\end{align}
For an isotropic material, the tensor of elastic moduli is given by
\begin{align}
\label{C}
C_{ijkl}=\lambda\, \delta_{ij}\delta_{kl}
+\mu\big(\delta_{ik}\delta_{jl}+\delta_{il}\delta_{jk})\,,
\end{align}
where $\lambda$ and $\mu$ are the Lam{\'e} constants.

In an unbounded medium and under the assumption of zero initial conditions,
which means that $\Bu(\rr,t_0)$ and $\dot\Bu(\rr,t_0)$ are zero 
for $t_0\rightarrow-\infty$, 
the solution of $\Bu$ reads
\begin{align}
u_i(\rr,t)=
\label{u-M}
\int_{-\infty}^t \int_{-\infty}^\infty
G_{ij}(\rr-\rr', t-t')\, F_{j}(\rr',t')\, \d \rr'\, \d t'\, .
\end{align}
When the material is isotropic and infinitely extended, 
the three-dimensional elastodynamic Green tensor reads~\citep{Love,Eringen75}
\begin{align}
\label{GT}
G_{ij}(\rr,t)&=\frac{1}{4\pi\rho }\, 
\Bigg\{
\frac{\delta_{ij}}{r c^2_\TT}\, \delta(t-r/c_\TT)
+\frac{x_i x_j}{r^3}\,
\bigg(
\frac{1}{c^2_\TL}\,  \delta(t-r/c_\TL)
-\frac{1}{c^2_\TT}\,  \delta(t-r/c_\TT)\bigg)\nonumber\\
&\hspace{15mm}
+\bigg(\frac{3x_i x_j}{r^2}-\delta_{ij}\bigg)
\frac{1}{r}\,\int_{1/c_\TL}^{1/c_\TT}\kappa\,\delta(t-\kappa r)\,\d \kappa
\Bigg\}\,,
\end{align}
where $r=\sqrt{x_1^2+x_2^2+x_3^2}$
and $\kappa$ is a dummy variable with the dimension
$1/[\text{velocity}]$.
Here $c_\TL$ and $c_\TT$ denote the velocities of the 
longitudinal and transversal elastic waves (sometimes called P- and S-waves). 
The two sound-velocities are given in terms of the Lam\'e constants ($c_\TT<c_\TL$)
\begin{align}
\label{c}
c_{\TL}=\sqrt{\frac{2\mu+\lambda}{\rho}}\,,\qquad
c_{\TT}=\sqrt{\frac{\mu}{\rho}}\, .
\end{align}

Substituting the Green tensor~(\ref{GT}) into Eq.~(\ref{u-M})
and integrating in time $t'$, 
the retarded displacement vector is obtained as (see also~\citep{Hudson})
\begin{align}
u_i(\rr,t)&=
\label{u-ret}
\frac{1}{4\pi\rho}\,\int_\VV 
\bigg\{
\frac{1}{c^2_\TT}\bigg(\frac{\delta_{ij}}{R}
-\frac{R_i R_j}{R^3}\bigg) F_{j}(\rr',t_\TT)
+\frac{1}{c^2_\TL}\,\frac{R_i R_j}{R^3}\,F_{j}(\rr',t_\TL)
\nonumber\\
&\hspace{16mm}
+\bigg(\frac{3R_i R_j}{R^3}-\frac{\delta_{ij}}{R}\bigg)
\int_{1/c_\TL}^{1/c_\TT} \kappa\, F_{j}(\rr',t_\kappa)\, \d \kappa
\bigg\}\,\d \rr'\, ,
\end{align}
where the so-called retarded times are given by
\begin{align}
\label{tT}
t_\TT&=t-\frac{R}{c_\TT}\, ,\\
\label{tL}
t_\TL&=t-\frac{R}{c_\TL}\, ,\\
\label{tkappa}
t_\kappa&=t-\kappa R\, .
\end{align}
Here 
$t_\TT$ and $t_\TL$ are the transversal retarded time and
the longitudinal retarded time, respectively.
The retarded time $t_\kappa$ is an effective retarded time for the 
$\kappa$-integration with the limits $(1/c_\TL,1/c_\TT)$.
Since $c_\TL>c_\TT$, the retarded times fulfill:
$t_\TT>t_\TL$ and $t_\kappa\in[t_\TL,t_\TT]$.
The idea of a retarded scalar potential was first developed 
by \citet{Lorenz61} 
in 1861 while studying waves in the theory of elasticity.
\citet{Love04} introduced the 
retarded potentials based on the Helmholtz decomposition 
(see also~\citep{Achenbach,Miklowitz}).
Our approach in this paper is more direct and straightforward, since
we have introduced the retarded displacement vector~(\ref{u-ret}) as 
the causal solution of the Navier equation~(\ref{Navier}).

The time-derivative of Eq.~(\ref{u-ret}) gives 
the retarded particle velocity vector
\begin{align}
v_{i}(\rr,t)&=
\label{v-ret}
\frac{1}{4\pi\rho}\,\int_\VV 
\bigg\{
\frac{1}{c^2_\TT}\bigg(\frac{\delta_{ij}}{R}
-\frac{R_i R_j}{R^3}\bigg) \pd_t F_{j}(\rr',t_\TT)
+\frac{1}{c^2_\TL}\,\frac{R_i R_j}{R^3}\,\pd_t F_{j}(\rr',t_\TL)
\nonumber\\
&\hspace{16mm}
+\bigg(\frac{3R_i R_j}{R^3}-\frac{\delta_{ij}}{R}\bigg)
\int_{1/c_\TL}^{1/c_\TT} \kappa\, \pd_t F_{j}(\rr',t_\kappa)\, \d \kappa
\bigg\}\,\d \rr'\, .
\end{align}
The gradient of Eq.~(\ref{u-ret}) and using the relation
\begin{align}
&\pd_k F_j(\rr',t-R/c)=-\frac{R_k}{c R}\, \pd_t
 F_j(\rr',t-R/c)\,,\qquad
c=c_\TT,c_\TL,1/\kappa\,,
\end{align}
lead to the retarded distortion tensor 
\begin{align}
&\beta_{ik}(\rr,t)=
\label{B-ret}
-\frac{1}{4\pi\rho}\, 
\int_\VV 
\bigg\{
\frac{1}{c^2_\TT}\bigg(
\frac{\delta_{ij} R_k +\delta_{jk} R_i +\delta_{ik} R_j}{R^3}
-\frac{3R_iR_jR_k}{R^5}\bigg)F_{j}(\rr',t_\TT)\nonumber\\
&\qquad
+\frac{1}{c^3_\TT}\bigg(\delta_{ij}-\frac{R_i R_j}{R^2}\bigg) 
\frac{R_k}{R^2}\,\pd_t F_{j}(\rr',t_\TT)
-\frac{1}{c^2_\TL}\bigg(
\frac{\delta_{jk} R_i +\delta_{ik} R_j}{R^3}
-\frac{3R_iR_jR_k}{R^5}\bigg)F_{j}(\rr',t_\TL)\nonumber\\
&\quad
+\frac{1}{c^3_\TL}\,\frac{R_i R_jR_k }{R^4}\, \pd_t F_{j}(\rr',t_\TL)
-\bigg(\frac{\delta_{ij} R_k +3\delta_{jk} R_i +3\delta_{ik} R_j}{R^3}
-\frac{9R_iR_jR_k}{R^5}\bigg)
\int_{1/c_\TL}^{1/c_\TT} \kappa\, F_{j}(\rr',t_\kappa)\, \d \kappa
\nonumber\\
&\hspace{16mm}
+\bigg(\frac{3R_i R_j}{R^2}-\delta_{ij}\bigg)\frac{R_k}{R^2}
\int_{1/c_\TL}^{1/c_\TT} \kappa^2\, \pd_t F_{j}(\rr',t_\kappa)\, \d \kappa
\bigg\}\,\d \rr'\, .
\end{align}
Since the integrals~(\ref{v-ret}) and (\ref{B-ret}) are evaluated at the retarded
times, they are called retarded elastic fields.
The sources $\BF$ and $\pd_t \BF$ at the position $\rr'$ 
depend on the retarded times.
Thus, the sources are retarded due to the retarded times.
Although Eqs.~(\ref{v-ret}) and (\ref{B-ret}) have some similarities with 
the Jefimenko equations in electrodynamics~(\ref{E-J}) and (\ref{B-J}),
the retarded fields in elastodynamics are more complicated than
the retarded fields in electrodynamics.

\subsection{Time-harmonic forces}
Here the retarded fields~(\ref{u-ret}), (\ref{v-ret}), and (\ref{B-ret})
are applied to the special case of time-harmonic forces.
The time-harmonic force vector reads
\begin{align}
\label{F-harm}
F_j(\rr,t)=\text{Re}\,\big\{F_j(\rr)\e^{-\ii\omega t}\big\} \,,
\end{align}
where $\omega$ denotes the angular frequency, $\ii=\sqrt{-1}$ and
Re indicates that the real part should be taken.
In Eqs.~(\ref{u-ret}), (\ref{v-ret}), and (\ref{B-ret}) the force enters at the
retarded times as
\begin{align}
\label{F-harm-ret}
F_j(\rr,t_c)=\text{Re}\,\big\{F_j(\rr)\e^{-\ii\omega(t-R/c)}\big\} \,,
\qquad c=c_\TT, c_\TL,1/\kappa\,.
\end{align}
Substituting Eq.~(\ref{F-harm-ret}) into Eq.~(\ref{u-ret}) and
using the integral
\begin{align}
\label{int1}
\int_{1/c_\TL}^{1/c_\TT}\kappa\, \e^{\ii\omega R \kappa}\, \d\kappa
=\frac{1}{\omega^2 R^2}\bigg[
\e^{\ii\omega R/c_\TT}\bigg(1-\frac{\ii\omega R}{c_\TT}\bigg)-
\e^{\ii\omega R/c_\TL}\bigg(1-\frac{\ii\omega R}{c_\TL}\bigg)\bigg]\,,
\end{align}
the displacement vector takes the form
\begin{align}
\label{u-harm}
u_i(\rr,t)&=\text{Re} \Bigg[
\frac{\e^{-\ii\omega t}}{4\pi\rho}\,
\int_\VV
\Bigg\{
\frac{1}{c^2_\TT}
\bigg(\frac{\delta_{ij}}{R}-\frac{R_iR_j}{R^3}\bigg)\e^{\ii \omega R/c_\TT}
+\frac{1}{c^2_\TL}\,
\frac{R_iR_j}{R^3}\,\e^{\ii \omega R/c_\TL}
\nonumber\\
&\quad
+\frac{1}{\omega^2 R^2}
\bigg(\frac{3R_i R_j}{R^3}-\frac{\delta_{ij}}{R}\bigg)
\bigg[\e^{\ii\omega R/c_\TT}\bigg(1-\frac{\ii\omega R}{c_\TT}\bigg)
-\e^{\ii\omega R/c_\TL}\bigg(1-\frac{\ii\omega R}{c_\TL}\bigg)\bigg]
\Bigg\} F_j(\rr')\, \d \rr'\Bigg] \,.
\end{align}
The first two terms behave like $1/R$ and the last term
gives $1/R^2$ and $1/R^3$ contributions. 
The velocity vector is given by: $v_i=-\text{Re}\,[\ii \omega u_i]$.
Now substituting Eq.~(\ref{F-harm-ret}) into Eq.~(\ref{B-ret}) and
using the integrals~(\ref{int1}) and
\begin{align}
\label{int2}
\int_{1/c_\TL}^{1/c_\TT}\kappa^2\, \e^{\ii\omega R \kappa}\, \d\kappa
=\frac{2\,\ii}{\omega^3 R^3}\bigg[
\e^{\ii\omega R/c_\TT}\bigg(1-\frac{\ii\omega R}{c_\TT}
-\frac{\omega^2 R^2}{2 c_\TT^2}\bigg)
-\e^{\ii\omega R/c_\TL}\bigg(1-\frac{\ii\omega R}{c_\TL}
-\frac{\omega^2 R^2}{2 c_\TL^2}\bigg)\bigg]\,,
\end{align}
the distortion tensor is obtained
\begin{align}
\label{B-harm}
\beta_{ik}(\rr,t)&=\text{Re} \Bigg[
-\frac{\e^{-\ii\omega t}}{4\pi\rho}\,
\int_\VV
\Bigg\{
\bigg(\frac{\delta_{ij}R_k+ \delta_{jk}R_i+\delta_{ik}R_j}{R^3}
-\frac{6R_iR_jR_k}{R^5}\bigg) 
\bigg(\frac{1}{c_\TT^2}\, \e^{\ii\omega R/c_\TT}- 
\frac{1}{c_\TL^2}\, \e^{\ii\omega R/c_\TL}\bigg)\nonumber\\
&\qquad
+\frac{\delta_{ij} R_k}{c^2_\TT R^3}\,\e^{\ii\omega R/c_\TT}
\bigg(1-\frac{\ii\omega R}{c_\TT}\bigg)
+\ii \omega\, \frac{R_iR_jR_k}{R^4}
\bigg(\frac{1}{c_\TT^3}\, \e^{\ii\omega R/c_\TT}- 
\frac{1}{c_\TL^3}\, \e^{\ii\omega R/c_\TL}\bigg)\nonumber\\
&\qquad
-\frac{3}{\omega^2 R^2}
\bigg(\frac{\delta_{ij}R_k+ \delta_{jk}R_i+\delta_{ik}R_j}{R^3}
-\frac{5 R_iR_jR_k}{R^5}\bigg)
\bigg[\e^{\ii\omega R/c_\TT}\bigg(1-\frac{\ii\omega R}{c_\TT}\bigg)
\nonumber\\
&\hspace{8cm}
-\e^{\ii\omega R/c_\TL}\bigg(1-\frac{\ii\omega R}{c_\TL}\bigg)\bigg]
\Bigg\} F_j(\rr')\, \d \rr'\Bigg] \,.
\end{align}
Eq.~(\ref{B-harm}) consists of $1/R$, $1/R^2$, $1/R^3$ and $1/R^4$ 
terms.

\subsection{A non-uniformly moving point force}

Now we consider the non-uniform motion of a point force
of total strength $\BQ(t)$ situated at the position $\Bs(t)$,
then the non-uniformly moving point force vector is given by
\begin{align}
\label{F}
F_{i}(\rr,t)&=Q_i(t) \, \delta(\rr-\Bs(t))\,
\qquad
\text{for}\ \rr\in\R^3,\ t\in\R\,.
\end{align}
Moreover, only subsonic source-speeds will be admitted ($|\BV|<c_\TT$).
Substitution of Eq.~(\ref{F}) in Eq.~(\ref{u-ret}) 
and integration in $t'$ lead to (see~\citep{Lazar12})
\begin{align}
u_i(\rr,t)&=
\label{u-LW0}
\frac{1}{4\pi\rho}\,\int_\VV 
\bigg\{
\frac{1}{c^2_\TT}\bigg(\frac{\delta_{ij}}{R}
-\frac{R_i R_j}{R^3}\bigg) Q_{j}(t_\TT)\,\delta(\rr'-\Bs(t_\TT))
+\frac{1}{c^2_\TL}\,\frac{R_i R_j}{R^3}\,Q_{j}(t_\TL)\,\delta(\rr'-\Bs(t_\TL))
\nonumber\\
&\hspace{16mm}
+\bigg(\frac{3R_i R_j}{R^3}-\frac{\delta_{ij}}{R}\bigg)
\int_{1/c_\TL}^{1/c_\TT} \kappa\, Q_{j}(t_\kappa)\, 
\delta(\rr'-\Bs(t_\kappa))\,\d \kappa
\bigg\}\,\d \rr'\, .
\end{align}

Like in Eqs.~(\ref{dz})--(\ref{int-J}) we can
carry out the integration in $\rr'$ in Eq.~(\ref{u-LW0})
and we find the explicit expression for the 
displacement field of a non-uniformly moving point force which we call the 
elastodynamic Li\'enard-Wiechert potential of a point force (see also~\citep{Lazar12})
\begin{align}
\label{u-LW}
u_i(\rr,t)&=
\frac{1}{4\pi\rho}
\Bigg\{
\frac{1}{c^2_\TT}
\bigg[\bigg(\delta_{ij}-\frac{R_i(t')R_j(t')}{R^2(t')}\bigg)
 \frac{Q_j(t')}{P_\TT(t')}\bigg]\bigg|_{t'=t_\TT}
+\frac{1}{c^2_\TL}\,
\bigg[\frac{R_i(t')R_j(t')}{R^2(t')}\,
 \frac{Q_j(t')}{P_\TL(t')}\bigg]\bigg|_{t'=t_\TL}
\nonumber\\
&\qquad\qquad\qquad
+\int_{1/c_\TL}^{1/c_\TT}\d \kappa \, \kappa
\bigg[\bigg(\frac{3R_i(t') R_j(t')}{R^2(t')}-\delta_{ij}\bigg)
\frac{Q_j(t')}{P_\kappa(t')}\bigg]\bigg|_{t'=t_\kappa}
\Bigg\}\,, 
\end{align}
with 
\begin{align}
\label{Pc}
P_c(t')=R(t')-V_m(t')R_m(t')/c \,, \qquad c=c_\TT,c_\TL,1/\kappa\,,
\end{align}
where $\BV(t')=\pd_{t'}\Bs(t')$ denotes the arbitrary 
velocity of the non-uniformly moving point force.
 The retarded times $t_c=t'(\rr,t)$ are given now as the solution of the condition
\begin{align}
\label{t-ret}
t-t'-|\rr-\Bs(t')|/c=0\,,\qquad{\text{with}}\qquad c=c_\TT,c_\TL,1/\kappa\,.
\end{align}

The substitution of Eq.~(\ref{F}) into the retarded elastic 
fields~(\ref{v-ret}) and (\ref{B-ret}) and after the $\rr'$-integration 
gives expressions for the elastic fields 
similar the Heaviside-Schott-Feynman equations
\begin{align}
\label{v-HF}
v_i(\rr,t)&=
\frac{1}{4\pi\rho}
\Bigg\{
\frac{1}{c^2_\TT}\,
\pd_t \bigg[\bigg(\delta_{ij}-\frac{R_i(t')R_j(t')}{R^2(t')}\bigg)
 \frac{Q_j(t')}{P_\TT(t')}\bigg]\bigg|_{t'=t_\TT}
+\frac{1}{c^2_\TL}\,
\pd_t \bigg[\frac{R_i(t')R_j(t')}{R^2(t')}\,
 \frac{Q_j(t')}{P_\TL(t')}\bigg]\bigg|_{t'=t_\TL}
\nonumber\\
&\qquad\qquad\qquad
+\int_{1/c_\TL}^{1/c_\TT}\d \kappa \, \kappa\,
\pd_t \bigg[\bigg(\frac{3R_i(t') R_j(t')}{R^2(t')}-\delta_{ij}\bigg)
\frac{Q_j(t')}{P_\kappa(t')}\bigg]\bigg|_{t'=t_\kappa}
\Bigg\}\,, 
\end{align}
and
\begin{align}
\label{b-HF}
\beta_{ik}(\rr,t)&=
-\frac{1}{4\pi\rho}
\bigg\{
\frac{1}{c^2_\TT}
\bigg[
\bigg(\frac{\delta_{ij} R_k(t') +\delta_{jk} R_i(t') +\delta_{ik} R_j(t')}{R^2(t')}
-\frac{3R_i(t')R_j(t')R_k(t')}{R^4(t')}\bigg)
\frac{Q_j(t')}{P_\TT(t')}\bigg]\bigg|_{t'=t_\TT}\nonumber\\
&\qquad\qquad
+\frac{1}{c^3_\TT}\,\pd_t 
\bigg[
\bigg(\delta_{ij}-\frac{R_i(t')R_j(t')}{R^2(t')}\bigg)\frac{R_k(t')}{R(t')}
 \frac{Q_j(t')}{P_\TT(t')})\bigg]\bigg|_{t'=t_\TT}
\nonumber\\
&\qquad\quad
-\frac{1}{c^2_\TL}
\bigg[
\bigg(\frac{\delta_{jk} R_i(t') +\delta_{ik} R_j(t')}{R^2(t')}
-\frac{3R_i(t')R_j(t')R_k(t')}{R^4(t')}\bigg) \frac{Q_j(t')}{P_\TL(t')}\bigg]\bigg|_{t'=t_\TL}
\nonumber\\
&\qquad\quad
+\frac{1}{c^3_\TL}\,\pd_t
\bigg[
\frac{R_i(t')R_j(t')R_k(t') }{R^3(t')}\,
 \frac{Q_j(t')}{P_\TL(t')}\bigg]\bigg|_{t'=t_\TL}
\nonumber\\
&\ 
-\int_{1/c_\TL}^{1/c_\TT}\d\kappa\, \kappa
\bigg[
\bigg(\frac{\delta_{ij} R_k(t') +3\delta_{jk} R_i(t') +3\delta_{ik} R_j(t')}{R^2(t')}
-\frac{9R_i(t')R_j(t')R_k(t')}{R^4(t')}\bigg)
\frac{Q_j(t') }{P_\kappa(t')}\bigg]\bigg|_{t'=t_\kappa}
\nonumber\\
&\qquad\quad
+\pd_t 
\int_{1/c_\TL}^{1/c_\TT}\d \kappa\, \kappa^2
\bigg[
\bigg(\frac{3R_i(t') R_j(t')}{R^2(t')}-\delta_{ij}\bigg)\frac{R_k(t')}{R(t')}
\frac{Q_j(t')}{P_\kappa(t')}
\bigg]\bigg|_{t'=t_\kappa}
\bigg\}\,.
\end{align}
Eqs.~(\ref{v-HF}) and (\ref{b-HF}) are similar in some sense 
to the Heaviside-Schott-Feynman equations~(\ref{E-HF}) and (\ref{B-HF}).
But of course, they have a more complicated tensor structure.

Using Eqs.~(\ref{dt-Q})--(\ref{dt-PRRR}), we obtain for
the time derivative of the displacement field~(\ref{u-LW})
or from (\ref{v-HF}) (see also~\citep{Lazar12})
\begin{align}
\label{v}
&v_{i}(\rr,t)=\frac{1}{4\pi\rho}
\Bigg\{
\frac{1}{c^2_\TT}\bigg[
\bigg(\delta_{ij}R(t')-\frac{R_i(t')R_j(t')}{R(t')}\bigg)
 \frac{\dot{Q}_j(t')}{P^2_\TT(t')}\bigg]\bigg|_{t'=t_\TT}
+\frac{1}{c^2_\TL}
\bigg[
\frac{R_i(t')R_j(t')}{R(t')}\,
 \frac{\dot{Q}_j(t')}{P^2_\TL(t')}\bigg]\bigg|_{t'=t_\TL}
\nonumber\\
&\qquad\qquad\qquad
+\int_{1/c_\TL}^{1/c_\TT}\d\kappa\, \kappa \bigg[
\bigg(\frac{3R_i(t') R_j(t')}{R(t')}-\delta_{ij} R(t')\bigg)
\frac{\dot{Q}_j(t')}{P^2_\kappa(t')}\bigg]\bigg|_{t'=t_\kappa}
\nonumber\\
&\quad
+\frac{1}{c^2_\TT}\bigg[
\bigg(\delta_{ij}-\frac{R_i(t')R_j(t')}{R^2(t')}\bigg) \frac{Q_j(t')}{P^3_\TT(t')}
\bigg(\big[\dot{V}_m(t') R_m(t')-V^2(t')\big]
\frac{R(t')}{c_\TT}+V_m(t') R_m(t')\bigg)
\nonumber\\
&\qquad\qquad
+\frac{Q_j(t')}{R(t') P^2_\TT(t')}\big(V_i(t') R_j(t')+V_j(t') R_i(t')\big)
-\frac{2 Q_j(t') R_i(t') R_j(t') V_m(t') R_m(t')}{R^3(t') P^2_\TT(t')}\bigg]\bigg|_{t'=t_\TT}
\nonumber\\
&\quad
+\frac{1}{c^2_\TL}\bigg[\frac{R_i(t')R_j(t')}{R^2(t')}
 \frac{Q_j(t')}{P^3_\TL(t')}
\bigg(\big[\dot{V}_m(t') R_m(t')-V^2(t')\big]
\frac{R(t')}{c_\TL}+V_m(t') R_m(t')\bigg)
\nonumber\\
&\qquad\qquad
-\frac{Q_j(t')}{R(t') P^2_\TL(t')}\big(V_i(t') R_j(t')+V_j(t') R_i(t')\big)
+\frac{2 Q_j(t') R_i(t') R_j(t') V_m(t') R_m(t')}{R^3(t') P^2_\TL(t')}\bigg]\bigg|_{t'=t_\TL}
\nonumber\\
&
+\int_{1/c_\TL}^{1/c_\TT}
\d \kappa\, \kappa
\bigg[\bigg(\frac{3R_i(t') R_j(t')}{R^2(t')}-\delta_{ij}\bigg)
\frac{Q_j(t')}{P^3_\kappa(t')}
\Big(\big[\dot{V}_m(t') R_m(t') -V^2(t')\big] \kappa R(t') +V_m(t') R_m(t')\Big)
\nonumber\\
&\qquad\qquad\
-\frac{3Q_j(t')}{R(t') P^2_\kappa(t')}\big(V_i(t') R_j(t')+V_j(t') R_i(t')\big)
+\frac{6 Q_j(t')R_i(t') R_j(t') V_m(t') R_m(t') }{R^3(t') P^2_\kappa(t')}\bigg]\bigg|_{t'=t_\kappa}
\Bigg\}\,.
\end{align}
This is the velocity field (particle velocity) produced 
by a non-uniformly moving
point force in the Li\'enard-Wiechert type representation.
If we carry out the time-derivatives in Eq.~(\ref{b-HF}) and after arranging 
terms, 
we obtain the Li\'enard-Wiechert type representation of the distortion tensor 
of a non-uniformly moving point force (see also~\citep{Lazar12})
\begin{align}
\label{b}
&\beta_{ik}(\rr,t)=-\frac{1}{4\pi\rho}
\Bigg\{
\frac{1}{c^3_\TT}
\bigg[
\bigg(\delta_{ij}-\frac{R_i(t')R_j(t')}{R^2(t')}\bigg)
 \frac{R_k(t') \dot{Q}_j(t')}{P^2_\TT(t')}\bigg]\bigg|_{t'=t_\TT}
+\frac{1}{c^3_\TL}
\bigg[
\frac{R_i(t')R_j(t')R_k(t')}{R^2(t')}\,
 \frac{\dot{Q}_j(t')}{P^2_\TL(t')}\bigg]\bigg|_{t'=t_\TL}
\nonumber\\
&\qquad\qquad\qquad\qquad
+\int_{1/c_\TL}^{1/c_\TT}\d \kappa\, \kappa^2
\bigg[\bigg(\frac{3R_i(t') R_j(t')}{R^2(t')}-\delta_{ij}\bigg)
\frac{R_k(t')\dot{Q}_j(t')}{P^2_\kappa(t')}\bigg]\bigg|_{t'=t_\kappa}
\nonumber\\
&
+\frac{1}{c^2_\TT}\bigg[
\bigg(\delta_{ij}-\frac{R_i(t')R_j(t')}{R^2(t')}\bigg)
 \frac{Q_j(t')}{P^3_\TT(t')}
\bigg(\frac{\dot{V}_m(t') R_m(t')}{c^2_\TT} R_k(t')
+\bigg(1-\frac{V^2(t')}{c^2_\TT}\bigg)R_k(t')
- \frac{P_\TT(t')}{c_\TT}\, V_k(t')\bigg)
\nonumber\\
&\qquad\quad
+\frac{Q_j(t')R_j(t')}{R^2(t') P_\TT(t')}
\bigg(\delta_{ik}+\frac{V_i(t')R_k(t')}{c_\TT P_\TT(t')}\bigg)
+\frac{Q_j(t')R_i(t')}{R^2(t') P_\TT(t')}
\bigg(\delta_{jk}+\frac{V_j(t')R_k(t')}{c_\TT P_\TT(t')}\bigg)
\nonumber\\
&\qquad\quad
-\frac{2 Q_j(t')R_i(t') R_j(t') R_k(t')}{R^3(t') P^2_\TT(t')}\bigg]\bigg|_{t'=t_\TT}
\nonumber\\
& 
+\frac{1}{c^2_\TL}\bigg[\frac{R_i(t')R_j(t')}{R^2(t')}\, 
\frac{Q_j(t')}{P^3_\TL(t')}
\bigg(\frac{\dot{V}_m(t') R_m(t')}{c^2_\TL}\, R_k(t')
+\bigg(1-\frac{V^2(t')}{c^2_\TL}\bigg)R_k(t')
- \frac{P_\TL(t')}{c_\TL}\, V_k(t')\bigg)
\nonumber\\
&\qquad\quad
-\frac{Q_j(t')R_j(t')}{R^2(t') P_\TL(t')}
\bigg(\delta_{ik}+\frac{V_i(t')R_k(t')}{c_\TL P_\TL(t')}\bigg)
-\frac{Q_j(t')R_i(t')}{R^2(t') P_\TL(t')}
\bigg(\delta_{jk}+\frac{V_j(t')R_k(t')}{c_\TL P_\TL(t')}\bigg)
\nonumber\\
&\qquad\quad
+\frac{2 Q_j(t')R_i(t') R_j(t') R_k(t')}{R^3(t') P^2_\TL(t')}\bigg]\bigg|_{t'=t_\TL}
\nonumber\\
& 
+\int_{1/c_\TL}^{1/c_\TT}
\d \kappa\, \kappa
\bigg[\bigg(\frac{3R_i(t') R_j(t')}{R^2(t')}-\delta_{ij}\bigg)
\frac{Q_j(t')}{P^3_\kappa(t')}
\bigg(\kappa^2 \dot{V}_m(t') R_m(t') R_k(t')
\nonumber\\
&\hspace{8cm}
+\big(1-\kappa^2 V^2(t')\big)R_k(t')
- \kappa\, P_\kappa(t')  V_k(t')\bigg)
\nonumber\\
&\qquad\quad 
-\frac{3 Q_j(t')R_j(t')}{R^2(t') P_\kappa(t')}
\bigg(\delta_{ik}+\frac{\kappa V_i(t')R_k(t')}{P_\kappa(t')}\bigg)
-\frac{3 Q_j(t')R_i(t')}{R^2(t') P_\kappa(t')}
\bigg(\delta_{jk}+\frac{\kappa V_j(t')R_k(t')}{P_\kappa(t')}\bigg)
\nonumber\\
&\qquad\quad
+\frac{6 Q_j(t')R_i(t') R_j(t') R_k(t')}{R^3(t') P^2_\kappa(t')}\bigg]\bigg|_{t'=t_\kappa}
\Bigg\}\,.
\end{align}
The fields~(\ref{v}) and (\ref{b}) 
consist of near fields which are the nonradiation  parts and
they fall off as $1/R^2$, and of far fields which are the radiation parts due to 
$\dot{\BQ}$ and $\dot{\BV}$ terms and they fall off as $1/R$.
Thus, there is a clear separation of $\Bbeta$ and $\Bv$ into two parts each, one 
which involves radiation and goes to zero for $\dot{\BV}=0$ and $\dot{q}=0$,
and one which involves only the velocity, $\BV$, and yields to the static
field for a point force having $\BV=0$:
\begin{align}
\beta_{ik}(\rr,t)&=\beta^{\text{nonr}}_{ik}(\rr,t)+\beta^{\text{rad}}_{ik}(\rr,t)\\
v_{i}(\rr,t)&=v^{\text{nonr}}_{i}(\rr,t)+v^{\text{rad}}_{i}(\rr,t)\,.
\end{align}
$\Bbeta^{\text{nonr}}$ and $\Bv^{\text{nonr}}$ are called the nonradiation 
or velocity-dependent fields and 
$\Bbeta^{\text{rad}}$ and $\Bv^{\text{rad}}$ are called the 
radiation fields or the acceleration fields.
Using a little algebra the nonradiation parts read
\begin{align}
\label{v-nonr}
&v^{\text{nonr}}_{i}(\rr,t)=\frac{1}{4\pi\rho}
\Bigg\{
\frac{1}{c^2_\TT}\bigg[
\bigg(\delta_{ij}-\frac{R_i(t')R_j(t')}{R^2(t')}\bigg) \frac{Q_j(t')}{P^3_\TT(t')}
\bigg(\Big(1-\frac{V^2(t')}{c_\TT^2}\Big) V_m(t') R_m(t')
-\frac{P_\TT(t')}{c_\TT}\, V^2(t')\bigg)
\nonumber\\
&\qquad
+\frac{Q_j(t')}{R(t') P^2_\TT(t')}\big(V_i(t') R_j(t')+V_j(t') R_i(t')\big)
-\frac{2 Q_j(t') R_i(t') R_j(t') V_m(t') R_m(t')}{R^3(t') P^2_\TT(t')}\bigg]\bigg|_{t'=t_\TT}
\nonumber\\
&\quad
+\frac{1}{c^2_\TL}\bigg[\frac{R_i(t')R_j(t')}{R^2(t')}
 \frac{Q_j(t')}{P^3_\TL(t')}
\bigg(
\Big(1-\frac{V^2(t')}{c_\TL^2}\Big) V_m(t') R_m(t')
-\frac{P_\TL(t')}{c_\TL}\, V^2(t')
\bigg)
\nonumber\\
&\qquad
-\frac{Q_j(t')}{R(t') P^2_\TL(t')}\big(V_i(t') R_j(t')+V_j(t') R_i(t')\big)
+\frac{2 Q_j(t') R_i(t') R_j(t') V_m(t') R_m(t')}{R^3(t') P^2_\TL(t')}\bigg]\bigg|_{t'=t_\TL}
\nonumber\\
&\quad
+\int_{1/c_\TL}^{1/c_\TT}
\d \kappa\, \kappa
\bigg[\bigg(\frac{3R_i(t') R_j(t')}{R^2(t')}-\delta_{ij}\bigg)
\frac{Q_j(t')}{P^3_\kappa(t')}
\Big(
\big(1-\kappa^2 V^2(t')\big) V_m(t') R_m(t')
- \kappa\, P_\kappa(t')  V^2(t')\Big)
\nonumber\\
&\qquad
-\frac{3Q_j(t')}{R(t') P^2_\kappa(t')}\big(V_i(t') R_j(t')+V_j(t') R_i(t')\big)
+\frac{6 Q_j(t')R_i(t') R_j(t') V_m(t') R_m(t') }{R^3(t') P^2_\kappa(t')}\bigg]\bigg|_{t'=t_\kappa}
\Bigg\}
\end{align}
and
\begin{align}
\label{b-nonr}
&\beta^{\text{nonr}}_{ik}(\rr,t)=-\frac{1}{4\pi\rho}
\Bigg\{
\frac{1}{c^2_\TT}\bigg[
\bigg(\delta_{ij}-\frac{R_i(t')R_j(t')}{R^2(t')}\bigg)
 \frac{Q_j(t')}{P^3_\TT(t')}
\bigg(
\Big(1-\frac{V^2(t')}{c^2_\TT}\Big)R_k(t')
- \frac{P_\TT(t')}{c_\TT}\, V_k(t')\bigg)
\nonumber\\
&\qquad\quad
+\frac{Q_j(t')R_j(t')}{R^2(t') P_\TT(t')}
\bigg(\delta_{ik}+\frac{V_i(t')R_k(t')}{c_\TT P_\TT(t')}\bigg)
+\frac{Q_j(t')R_i(t')}{R^2(t') P_\TT(t')}
\bigg(\delta_{jk}+\frac{V_j(t')R_k(t')}{c_\TT P_\TT(t')}\bigg)
\nonumber\\
&\qquad\quad
-\frac{2 Q_j(t')R_i(t') R_j(t') R_k(t')}{R^3(t') P^2_\TT(t')}\bigg]\bigg|_{t'=t_\TT}
\nonumber\\
&\qquad 
+\frac{1}{c^2_\TL}\bigg[\frac{R_i(t')R_j(t')}{R^2(t')}\, 
\frac{Q_j(t')}{P^3_\TL(t')}
\bigg(
\Big(1-\frac{V^2(t')}{c^2_\TL}\Big)R_k(t')
- \frac{P_\TL(t')}{c_\TL}\, V_k(t')\bigg)
\nonumber\\
&\qquad\quad
-\frac{Q_j(t')R_j(t')}{R^2(t') P_\TL(t')}
\bigg(\delta_{ik}+\frac{V_i(t')R_k(t')}{c_\TL P_\TL(t')}\bigg)
-\frac{Q_j(t')R_i(t')}{R^2(t') P_\TL(t')}
\bigg(\delta_{jk}+\frac{V_j(t')R_k(t')}{c_\TL P_\TL(t')}\bigg)
\nonumber\\
&\qquad\quad
+\frac{2 Q_j(t')R_i(t') R_j(t') R_k(t')}{R^3(t') P^2_\TL(t')}\bigg]\bigg|_{t'=t_\TL}
\nonumber\\
&\qquad 
+\int_{1/c_\TL}^{1/c_\TT}
\d \kappa\, \kappa
\bigg[\bigg(\frac{3R_i(t') R_j(t')}{R^2(t')}-\delta_{ij}\bigg)
\frac{Q_j(t')}{P^3_\kappa(t')}
\Big(
\big(1-\kappa^2 V^2(t')\big)R_k(t')
- \kappa\, P_\kappa(t')  V_k(t')\Big)
\nonumber\\
&\qquad\quad 
-\frac{3 Q_j(t')R_j(t')}{R^2(t') P_\kappa(t')}
\bigg(\delta_{ik}+\frac{\kappa V_i(t')R_k(t')}{P_\kappa(t')}\bigg)
-\frac{3 Q_j(t')R_i(t')}{R^2(t') P_\kappa(t')}
\bigg(\delta_{jk}+\frac{\kappa V_j(t')R_k(t')}{P_\kappa(t')}\bigg)
\nonumber\\
&\qquad\quad
+\frac{6 Q_j(t')R_i(t') R_j(t') R_k(t')}{R^3(t') P^2_\kappa(t')}\bigg]\bigg|_{t'=t_\kappa}
\Bigg\}\,.
\end{align}
Using the relation~(\ref{Pc}), one verifies
\begin{align}
v^{\text{nonr}}_{i}(\rr,t)=-\Big[\beta^{\text{nonr}}_{ik}(\rr,t')\,V_k(t')\Big]\Big|_{t'=\tr}\,,\qquad{\text{for}}\qquad \tr=t_\TT,t_\TL,t_\kappa\,. 
\end{align}
On the other hand, the elastodynamic radiation fields are given by
\begin{align}
\label{v-rad}
v^{\text{rad}}_{i}(\rr,t)&=\frac{1}{4\pi\rho}
\Bigg\{
\frac{1}{c^2_\TT}\bigg[
\bigg(\delta_{ij} R(t')-\frac{R_i(t')R_j(t')}{R(t')}\bigg)
\bigg( \frac{\dot{Q}_j(t')}{P^2_\TT(t')}
+\frac{Q_j(t')\dot{V}_m(t') R_m(t')}{c_\TT\, P^3_\TT(t')}\bigg)
\bigg]\bigg|_{t'=t_\TT}
\nonumber\\
&\qquad\quad
+\frac{1}{c^2_\TL}
\bigg[
\frac{R_i(t')R_j(t')}{R(t')}
\bigg( \frac{\dot{Q}_j(t')}{P^2_\TL(t')}
+\frac{Q_j(t')\dot{V}_m(t') R_m(t')}{c_\TL\, P^3_\TL(t')}\bigg)
\bigg]\bigg|_{t'=t_\TL}
\nonumber\\
&\quad
+\int_{1/c_\TL}^{1/c_\TT}\d\kappa\, \kappa \bigg[
\bigg(\frac{3R_i(t') R_j(t')}{R(t')}-\delta_{ij} R(t')\bigg)
\bigg(\frac{\dot{Q}_j(t')}{P^2_\kappa(t')}
+\frac{\kappa\, Q_j(t')\dot{V}_m(t') R_m(t')}{P^3_\kappa(t')}\bigg)
\bigg]\bigg|_{t'=t_\kappa}\Bigg\}
\end{align}
and
\begin{align}
\label{b-rad}
&\beta^{\text{rad}}_{ik}(\rr,t)=-\frac{1}{4\pi\rho}
\Bigg\{
\frac{1}{c^3_\TT}
\bigg[
\bigg(\delta_{ij} R_k(t')-\frac{R_i(t')R_j(t')R_k(t')}{R^2(t')}\bigg) 
\bigg( \frac{\dot{Q}_j(t')}{P^2_\TT(t')}
+\frac{Q_j(t')\dot{V}_m(t') R_m(t')}{c_\TT\, P^3_\TT(t')}\bigg)
\bigg]\bigg|_{t'=t_\TT}
\nonumber\\
&\qquad\qquad\qquad
+\frac{1}{c^3_\TL}
\bigg[\frac{R_i(t')R_j(t')R_k(t')}{R^2(t')}
\bigg( \frac{\dot{Q}_j(t')}{P^2_\TL(t')}
+\frac{Q_j(t')\dot{V}_m(t') R_m(t')}{c_\TL\, P^3_\TL(t')}\bigg)
\bigg]\bigg|_{t'=t_\TL}
\nonumber\\
&\quad
+\int_{1/c_\TL}^{1/c_\TT}\d \kappa\, \kappa^2
\bigg[\bigg(\frac{3R_i(t') R_j(t')R_k(t')}{R^2(t')}-\delta_{ij}R_k(t')\bigg)
\bigg(\frac{\dot{Q}_j(t')}{P^2_\kappa(t')}
+\frac{\kappa\, Q_j(t')\dot{V}_m(t') R_m(t')}{P^3_\kappa(t')}\bigg)
\bigg]\bigg|_{t'=t_\kappa}\Bigg\}\,.
\end{align}
The following relation holds for the radiation parts
\begin{align}
v^{\text{rad}}_{i}(\rr,t)=-\Big[\beta^{\text{rad}}_{ik}(\rr,t')\,R_k(t')/c R(t')\Big]\Big|_{t'=\tr}\,,\qquad{\text{for}}\qquad c=c_\TT,c_\TL,1/\kappa\,. 
\end{align}
The elastodynamic radiation is caused by the time-change of the 
magnitude of the point force $\dot{q}$ and the acceleration $\dot\BV$. 
\\
{\bf Remark:}
Some remarks concerning transonic and supersonic motions will be given. 
If the velocity of the moving point force is $c_\TT<|\BV|<c_\TL$, then 
the motion is transonic. Since the source of the elastodynamic fields travels 
faster than the elastodynamic waves, some sort of shock front is set up. 
The disturbances initiated at each point of its
track by the moving point force arrive simultaneously at the  
surface of a right circular cone whose vertex is at the point force.
The semi-vertical angle of this cone is: $\theta_\TT=\sin^{-1} (c_\TT/|\BV|)$.
This cone is the Mach cone with respect to $c_\TT$.
The field depending on $c_\TT$  at any point is zero outside this Mach
cone. 
If the velocity of the moving point force is $c_\TL<|\BV|$, then 
the motion is supersonic. Another shock front and Mach cone 
with respect to $c_\TL$ are built. 
The semi-vertical angle of this Mach cone is: 
$\theta_\TL=\sin^{-1} (c_\TL/|\BV|)$.
Outside the Mach cones the elastodynamic fields are zero 
and inside the elastodynamic fields can be obtained from the 
Li\'enard-Wiechert potentials with modified retarded times. 
For the transonic and supersonic motions the retarded times 
$t_\TT$ and $t_\TL$, respectively, have more than one solution
(at least two solutions), each. 
Depending on the specific motion, any number of retarded times,
not just one transversal retarded time  and one longitudinal retarded time,
may be associated with a given time $t$.
For points outside the Mach cone no retarded time exists and the fields must
be zero. For points that are inside the Mach cone many retarded 
times may exist (see, e.g.,~\citep{CD,Smith}). 
Also  we want to mention that the non-uniform motion of a supersonic screw
dislocation has been analyzed in~\citep{XM80}.

\subsection{The Stokes solution as limit of a non-uniformly moving point force}
If the position of the point force is fixed that means that 
$\Bs$ is time-independent and therefore $\BV=0$, 
we recover from the displacement~(\ref{u-LW})
the famous Stokes solution~\cite{Stokes} 
of a concentrated point force with time-dependent
magnitude (e.g.~\citep{Eringen75,WS,Gurtin})
\begin{align}
\label{u-St}
u_i(\rr,t)&=
\frac{1}{4\pi\rho R}\,
\Bigg\{
\frac{1}{c^2_\TT}
\bigg(\delta_{ij}-\frac{R_iR_j}{R^2}\bigg)Q_j(t-R/c_\TT)
+\frac{1}{c^2_\TL}\,
\frac{R_iR_j}{R^2}\,Q_j(t-R/c_\TL)
\nonumber\\
&\qquad
+\bigg(\frac{3R_i R_j}{R^2}-\delta_{ij}\bigg)
\int_{1/c_\TL}^{1/c_\TT}
\kappa\, Q_j(t-\kappa R)\,\d \kappa\Bigg\} \,.
\end{align}
The first terms in Eq.~(\ref{u-St}) 
are usually called the far-field terms since they behave as $1/R$
and the last term in Eq.~(\ref{u-St}) 
is called the near-field term.
From Eq.~(\ref{v}), we obtain the particle velocity vector
\begin{align}
\label{v-St}
v_i(\rr,t)&=
\frac{1}{4\pi\rho R}\,
\Bigg\{
\frac{1}{c^2_\TT}
\bigg(\delta_{ij}-\frac{R_iR_j}{R^2}\bigg)\dot{Q}_j(t-R/c_\TT)
+\frac{1}{c^2_\TL}\,
\frac{R_iR_j}{R^2}\,\dot{Q}_j(t-R/c_\TL)
\nonumber\\
&\qquad
+\bigg(\frac{3R_i R_j}{R^2}-\delta_{ij}\bigg)
\int_{1/c_\TL}^{1/c_\TT}
\kappa\, \dot{Q}_j(t-\kappa R)\,\d \kappa\Bigg\} \,.
\end{align}
From Eq.~(\ref{b}) and after some mathematical manipulations 
we find the corresponding displacement gradient
of the Stokes solution (e.g.~\citep{Eringen75,WS,Gurtin})
\begin{align}
\label{B-St}
\beta_{ik}(\rr,t)&=
-\frac{1}{4\pi\rho}
\Bigg\{
3\bigg(\frac{5 R_iR_jR_k}{R^5}
-\frac{\delta_{ij}R_k+ \delta_{jk}R_i+\delta_{ik}R_j}{R^3}\bigg)  
\int_{1/c_\TL}^{1/c_\TT}
\kappa\, Q_j(t-\kappa R)\,\d \kappa\nonumber\\
&\quad
+\bigg(\frac{6R_iR_jR_k}{R^5}
-\frac{\delta_{ij}R_k+ \delta_{jk}R_i+\delta_{ik}R_j}{R^3}\bigg)  
\bigg[
\frac{1}{c^2_\TL}\,Q_j(t-R/c_\TL)
-\frac{1}{c^2_\TT}\,Q_j(t-R/c_\TT)\bigg]\nonumber\\
&\qquad
+\frac{\delta_{ij} R_k}{c^2_\TT R^3}\, 
\bigg[Q_j(t-R/c_\TT)
+\frac{R}{c_\TT}\,\dot{Q}_j(t-R/c_\TT)\bigg]\nonumber\\
&\qquad
+\frac{R_iR_jR_k}{R^4}
\bigg[
\frac{1}{c^3_\TL}\,\dot{Q}_j(t-R/c_\TL)
-\frac{1}{c^3_\TT}\,\dot{Q}_j(t-R/c_\TT)\bigg]
\Bigg\}
 \,.
\end{align}
Thus, Eqs.~(\ref{u-LW}) and (\ref{b}) give the correct Stokes solution
as limit.
It can be seen that the $\dot{\BQ}$-terms in Eq.~(\ref{B-St})
behave as $1/R$ (far-field terms or radiation terms) and
the $\BQ$-terms behave as $1/R^2$ (near-field terms).
Also, it can be seen that the third line in Eq.~(\ref{B-St})
is analogous to Eq.~(\ref{E-q}).
Thus, the solutions~(\ref{u-LW}), (\ref{v}) and (\ref{b})
are the generalization of the Stokes solution toward the non-uniform
motion.
\\
{\bf Remark:}
\citet{BS} mentioned that \citet{Stokes} had conceived with
his solution the first mathematical model of an earthquake.
In this sense our solution of a non-uniformly point force may serve 
as a mathematical model of the non-uniform motion of sources (senders)
of elastodynamic waves (P- and S-seismic waves) in seismology.

The same class of problems considered in this paper for elastic waves in an
unbounded medium might be more significant, but also more complicated, for
point sources moving over the surface of a half-space and
the interaction with external or coupled fields (see, e.g., \citep{Achenbach}).  

\subsection{Time-harmonic point force}
Here we consider the special case in which the body force is harmonic in time.
For a time-harmonic point force with 
the spatial part
\begin{align}
\label{F-harm2}
F_j(\rr)=Q_j\,\delta(\rr-\rr')\,,
\end{align}
Eq.~(\ref{u-harm}) gives directly
\begin{align}
\label{u-harm2}
u_i(\rr,t)&=\text{Re} \Bigg[
\frac{Q_j\, \e^{-\ii\omega t}}{4\pi\rho}\,
\Bigg\{
\frac{1}{c^2_\TT}
\bigg(\frac{\delta_{ij}}{R}-\frac{R_iR_j}{R^3}\bigg)\e^{\ii \omega R/c_\TT}
+\frac{1}{c^2_\TL}\,
\frac{R_iR_j}{R^3}\,\e^{\ii \omega R/c_\TL}
\nonumber\\
&\quad
+\frac{1}{\omega^2 R^2}
\bigg(\frac{3R_i R_j}{R^3}-\frac{\delta_{ij}}{R}\bigg)
\bigg[\e^{\ii\omega R/c_\TT}\bigg(1-\frac{\ii\omega R}{c_\TT}\bigg)
-\e^{\ii\omega R/c_\TL}\bigg(1-\frac{\ii\omega R}{c_\TL}\bigg)\bigg]
\Bigg\}\Bigg]\,,
\end{align}
which is in agreement with the expression given in \citep{Eringen75,Hudson}.
Moreover, the velocity vector reads $v_i=-\text{Re}\,[\ii \omega u_i]$,
and Eq.~(\ref{B-harm}) gives the elastic distortion
\begin{align}
\label{B-harm2}
\beta_{ik}(\rr,t)&=\text{Re} \Bigg[
-\frac{Q_j\, \e^{-\ii\omega t}}{4\pi\rho}\,
\Bigg\{
\bigg(\frac{\delta_{ij}R_k+ \delta_{jk}R_i+\delta_{ik}R_j}{R^3}
-\frac{6R_iR_jR_k}{R^5}\bigg) 
\bigg(\frac{1}{c_\TT^2}\, \e^{\ii\omega R/c_\TT}- 
\frac{1}{c_\TL^2}\, \e^{\ii\omega R/c_\TL}\bigg)\nonumber\\
&\qquad
+\frac{\delta_{ij} R_k}{c^2_\TT R^3}\,\e^{\ii\omega R/c_\TT}
\bigg(1-\frac{\ii\omega R}{c_\TT}\bigg)
+\ii \omega\, \frac{R_iR_jR_k}{R^4}
\bigg(\frac{1}{c_\TT^3}\, \e^{\ii\omega R/c_\TT}- 
\frac{1}{c_\TL^3}\, \e^{\ii\omega R/c_\TL}\bigg)\nonumber\\
&\qquad
-\frac{3}{\omega^2 R^2}
\bigg(\frac{\delta_{ij}R_k+ \delta_{jk}R_i+\delta_{ik}R_j}{R^3}
-\frac{5 R_iR_jR_k}{R^5}\bigg)
\bigg[\e^{\ii\omega R/c_\TT}\bigg(1-\frac{\ii\omega R}{c_\TT}\bigg)
\nonumber\\
&\hspace{8cm}
-\e^{\ii\omega R/c_\TL}\bigg(1-\frac{\ii\omega R}{c_\TL}\bigg)\bigg]
\Bigg\} \Bigg] \,.
\end{align}

\section*{Acknowledgement}
The author gratefully acknowledges the grants from the 
Deutsche Forschungsgemeinschaft (Grant Nos. La1974/2-1, La1974/3-1). 

\begin{appendix}
\setcounter{equation}{0}
\renewcommand{\theequation}{\thesection.\arabic{equation}}
\section{Appendix: Derivatives at the retarded time}

\label{appendix}

Here we give some useful relations of derivatives of quantities 
depending on the retarded time, which is the unique solution of the 
relation
\begin{align}
t-\tr-|\rr-\Bs(\tr)|/c=0\,.
\end{align}
First of all, we carry out the time derivatives, which are not
trivial because of the subtle relation between present and retarded time
(see also~\citep{Barton}):
\begin{align}
\label{dt-t}
 \bigg[\frac{\pd t'}{\pd t}\bigg]\bigg|_{t'=\tr}
=\bigg[\frac{R(t')}{P(t')}\bigg]\bigg|_{t'=\tr}
\end{align}
\begin{align}
\label{dt-Q}
\pd_t \big[Q_j(t')\big]\big|_{t'=\tr}
=\bigg[\frac{\pd t'}{\pd t}\, \frac{\pd Q_j(t')}{\pd t'}\bigg]\bigg|_{t'=\tr}
=\bigg[\frac{R(t')}{P(t')}\, \dot{Q}_j(t')\bigg]\bigg|_{t'=\tr}
\end{align}
\begin{align}
\label{dt-P}
\pd_t \bigg[\frac{1}{P(t')}\bigg]\bigg|_{t'=\tr}
=\bigg[\frac{1}{P^3(t')}\bigg(
\Big(\dot{V}_m(t') R_m(t') -V^2(t')\Big)\, 
\frac{R(t')}{c}+V_m(t') R_m(t')\bigg)\bigg]\bigg|_{t'=\tr}
\end{align}
\begin{align}
\label{dt-PR}
\pd_t \bigg[\frac{R_k(t')}{R(t')P(t')}\bigg]\bigg|_{t'=\tr}
&=\bigg[\frac{R_k(t')}{R(t')P^3(t')}\bigg(
\Big(\dot{V}_m(t') R_m(t') -V^2(t')\Big)\, \frac{R(t')}{c}
+V_m(t') R_m(t')\bigg)
\nonumber\\
&\qquad 
+\frac{1}{P^2(t')}\bigg(
V_m(t') R_m(t')\, \frac{R_k(t')}{R^2(t')}-V_k(t')\bigg)\bigg]\bigg|_{t'=\tr}\,
\end{align}
\begin{align}
\label{dt-PRR}
\pd_t \bigg[\frac{R_i(t')R_j(t')}{R^2(t')P(t')}\bigg]\bigg|_{t'=\tr}
&=\bigg[\frac{R_i(t')R_j(t')}{R^2(t')P^3(t')}\bigg(
\Big(\dot{V}_m(t') R_m(t') -V^2(t')\Big)\, \frac{R(t')}{c}
+V_m(t') R_m(t')\bigg)
\nonumber\\
&
-\frac{1}{R(t')P^2(t')}\big(
V_i(t') R_j(t')+V_j(t') R_i(t')\big)
+\frac{2R_i(t')R_j(t')V_m(t') R_m(t')}{R^3(t')P^2(t')}
\bigg]\bigg|_{t'=\tr}\,
\end{align}
\begin{align}
\label{dt-PRRR}
&\pd_t \bigg[\frac{R_i(t')R_j(t')R_k(t')}{R^3(t')P(t')}\bigg]\bigg|_{t'=\tr}
=\bigg[\frac{R_i(t')R_j(t')R_k(t')}{R^3(t')P^3(t')}\bigg(
\Big(\dot{V}_m(t') R_m(t') -V^2(t')\Big)\, \frac{R(t')}{c}
+V_m(t') R_m(t')\bigg)
\nonumber\\
&\hspace{3cm}
-\frac{1}{R^2(t')P^2(t')}\big(
V_i(t') R_j(t')R_k(t')+V_j(t') R_k(t') R_i(t')+V_k(t') R_i(t')R_j(t')\big)
\nonumber\\
&\hspace{3cm}
+\frac{3R_i(t')R_j(t') R_k(t')V_m(t') R_m(t')}{R^4(t')P^2(t')}
\bigg]\bigg|_{t'=\tr}\,.
\end{align}

\end{appendix}

\end{document}